\PassOptionsToPackage{unicode}{hyperref}
\PassOptionsToPackage{hyphens}{url}
\documentclass[
]{article}
\title{Relaxation of the parameter independence assumption in the
\texttt{bootComb} R package}
\author{Marc Y. R. Henrion\textsuperscript{1,2}}
\date{}

\usepackage{amsmath,amssymb}
\usepackage{lmodern}
\usepackage{iftex}
\ifPDFTeX
  \usepackage[T1]{fontenc}
  \usepackage[utf8]{inputenc}
  \usepackage{textcomp} 
\else 
  \usepackage{unicode-math}
  \defaultfontfeatures{Scale=MatchLowercase}
  \defaultfontfeatures[\rmfamily]{Ligatures=TeX,Scale=1}
\fi
\IfFileExists{upquote.sty}{\usepackage{upquote}}{}
\IfFileExists{microtype.sty}{
  \usepackage[]{microtype}
  \UseMicrotypeSet[protrusion]{basicmath} 
}{}
\makeatletter
\@ifundefined{KOMAClassName}{
  \IfFileExists{parskip.sty}{%
    \usepackage{parskip}
  }{
    \setlength{\parindent}{0pt}
    \setlength{\parskip}{6pt plus 2pt minus 1pt}}
}{
  \KOMAoptions{parskip=half}}
\makeatother
\usepackage{xcolor}
\IfFileExists{xurl.sty}{\usepackage{xurl}}{} 
\IfFileExists{bookmark.sty}{\usepackage{bookmark}}{\usepackage{hyperref}}
\hypersetup{
  pdftitle={Relaxation of the parameter independence assumption in the bootComb R package},
  pdfauthor={Marc Y. R. Henrion1,2},
  hidelinks,
  pdfcreator={LaTeX via pandoc}}
\usepackage[margin=1in]{geometry}
\usepackage{color}
\usepackage{fancyvrb}

\DefineVerbatimEnvironment{Highlighting}{Verbatim}{commandchars=\\\{\}}
\usepackage{framed}
\definecolor{shadecolor}{RGB}{248,248,248}
\newenvironment{Shaded}{\begin{snugshade}}{\end{snugshade}}

\newcommand{\AttributeTok}[1]{\textcolor[rgb]{0.77,0.63,0.00}{#1}}

\newcommand{\ConstantTok}[1]{\textcolor[rgb]{0.00,0.00,0.00}{#1}}
\newcommand{\ControlFlowTok}[1]{\textcolor[rgb]{0.13,0.29,0.53}{\textbf{#1}}}

\newcommand{\DecValTok}[1]{\textcolor[rgb]{0.00,0.00,0.81}{#1}}

\newcommand{\FloatTok}[1]{\textcolor[rgb]{0.00,0.00,0.81}{#1}}
\newcommand{\FunctionTok}[1]{\textcolor[rgb]{0.00,0.00,0.00}{#1}}

\newcommand{\NormalTok}[1]{#1}

\newcommand{\OtherTok}[1]{\textcolor[rgb]{0.56,0.35,0.01}{#1}}

\newcommand{\SpecialCharTok}[1]{\textcolor[rgb]{0.00,0.00,0.00}{#1}}

\newcommand{\StringTok}[1]{\textcolor[rgb]{0.31,0.60,0.02}{#1}}

\usepackage{graphicx}
\makeatletter
\def\maxwidth{\ifdim\Gin@nat@width>\linewidth\linewidth\else\Gin@nat@width\fi}
\def\maxheight{\ifdim\Gin@nat@height>\textheight\textheight\else\Gin@nat@height\fi}
\makeatother
\setkeys{Gin}{width=\maxwidth,height=\maxheight,keepaspectratio}
\makeatletter
\def\fps@figure{htbp}
\makeatother
\providecommand{\tightlist}{%
  \setlength{\itemsep}{0pt}\setlength{\parskip}{0pt}}
\newlength{\cslhangindent}
\newlength{\csllabelwidth}
\newlength{\cslentryspacingunit} 
\newenvironment{CSLReferences}[2] 
 {
  \setlength{\parindent}{0pt}
  \ifodd #1
  \let\oldpar\par
  \def\par{\hangindent=\cslhangindent\oldpar}
  \fi
  \setlength{\parskip}{#2\cslentryspacingunit}
 }%
 {}
\usepackage{calc}

\usepackage{setspace}
\ifLuaTeX
  \usepackage{selnolig}  
\fi

\begin{document}
\maketitle

\textsuperscript{1} Malawi - Liverpool - Wellcome Clinical Research
Programme, Blantyre, Malawi

\textsuperscript{2} Liverpool School of Tropical Medicine, Liverpool, UK

\hypertarget{key-words}{%
\section*{Key words}\label{key-words}}
\addcontentsline{toc}{section}{Key words}

Biostatistics, R, confidence intervals, bootstrap, estimation

\hypertarget{word-count}{%
\section*{Word count}\label{word-count}}
\addcontentsline{toc}{section}{Word count}

\textbf{Abstract}: 142 words

\textbf{Main text} (excluding abstract, key features, references): 1,518
words

\newpage

\hypertarget{abstract}{%
\section*{Abstract}\label{abstract}}
\addcontentsline{toc}{section}{Abstract}

\textbf{Background} The \texttt{bootComb} R package allows researchers
to derive confidence intervals with correct target coverage for
arbitrary combinations of arbitrary numbers of independently estimated
parameters. Previous versions (\textless{} 1.1.0) of \texttt{bootComb}
used independent bootstrap sampling and required that the parameters
themselves are independent - an unrealistic assumption in some
real-world applications.

\textbf{Findings} Using Gaussian copulas to define the dependence
between parameters, the \texttt{bootComb} package has been extended to
allow for dependent parameters.

\textbf{Implications} The updated \texttt{bootComb} package can now
handle cases of dependent parameters, with users specifying a
correlation matrix defining the dependence structure. While in practice
it may be difficult to know the exact dependence structure between
parameters, \texttt{bootComb} allows running sensitivity analyses to
assess the impact of parameter dependence on the resulting confidence
interval for the combined parameter.

\textbf{Availability} \texttt{bootComb} is available from the
Comprehensive R Archive Network
(\url{https://CRAN.R-project.org/package=bootComb}).

\hypertarget{introduction}{%
\section*{Introduction}\label{introduction}}
\addcontentsline{toc}{section}{Introduction}

The \texttt{bootcomb} R package Henrion (2021) was recently published.
This package for the statistical computation environment R (R Core Team,
2021) allows researchers to derive confidence intervals with correct
coverage for combinations of independently estimated parameters.
Important applications include adjusting a prevalence for estimated test
sensitivity and specificity (e.g. Mandolo et al. (2021)) or combining
conditional prevalence estimates (e.g. Stockdale et al. (2020)).

Briefly, for each of the input parameters, \texttt{bootComb} finds a
best-fit parametric distribution based on the confidence interval for
that parameter estimate. \texttt{bootComb} then uses the parametric
bootstrap to sample many sets of parameter estimates from these best-fit
distributions and computes the corresponding combined parameter estimate
for each set. This builds up an empirical distribution of parameter
estimates for the combined parameter. Finally, \texttt{bootComb} uses
either the percentile or the highest density interval method to derive a
confidence interval for the combined parameter estimate. Full details of
the algorithm are given in Henrion (2021).

A key point of the algorithm is that the best-fit distributions for the
different parameters are sampled from independently. This requires the
parameters to be independent. This may not be a realistic assumption in
some real-world applications.

While for most practical applications the input parameters are typically
estimated from independent experiments (otherwise the combined parameter
could be directly estimated), the parameters themselves may not be
independent. This is for instance the case when adjusting a prevalence
for the diagnostic test's sensitivity and specificity. The latter two
parameters are not independent: higher sensitivity can be achieved by
lowering specificity and vice versa.

If the experiments estimating these parameters are sufficiently large,
then the violation of the assumption of parameter independence may only
have negligible impact on the resulting confidence interval for the
combined parameter. However, for the sake of general applicability and
to allow running sensitivity analyses, the author felt it was beneficial
to extend \texttt{bootComb} to handle dependent parameters.

\hypertarget{methods}{%
\section*{Methods}\label{methods}}
\addcontentsline{toc}{section}{Methods}

Copulas are multivariate distribution functions where the marginal
probability distribution of each variable is the uniform distribution on
the interval \([0,1]\). Copulas allow to specify the intercorrelation
between random variables. An important probability theory result,
Sklar's Theorm (Sklar, 1959), states that any multivariate probability
distribution can be expressed in terms of its univariate marginal
distributions and a copula defining the dependence between the
variables.

Mathematically, let \(X_1,X_2\ldots,X_d\) be \(d\) random variables and
define \(U_i = F_i(X_i), i=1,\ldots,d\). Then the copula \(C\) of
\((X_1,\ldots,X_d)\) is defined as the joint cumulative distribution
function of \((U_1,\ldots,U_d)\):

\[
C(u_1,\ldots,u_d)=Pr(U_1\leq u_1,\ldots,U_d\leq u_d)
\]

Assume that the marginal distributions,
\(F_i(x)=Pr[X_i\leq x], i=1,\ldots,d\) are continuous. Then, via the
probability integral transform (Angus, 1994), the random vector
\((U_1,U_2,\ldots,U_d)\) has marginals that are uniformly distributed on
\([0,1]\).

\texttt{bootComb} makes use of the fact that the above can be reversed:
given a sample \((u_1,\ldots,u_d)\), a sample for \((X_1,\ldots,X_d)\)
can be obtained by
\((x_1,\ldots,x_d)=(F^{-1}_1(u_1),\ldots,F^{-1}_d(u_d))\). The inverse
functions \(F^{-1}_i(u)\) will be defined if the marginals \(F_i(x)\)
are continuous. For the use of \texttt{bootComb}, where users input
confidence intervals for an estimated numeric parameter, this will
always be the case.

\texttt{bootComb} will proceed as follows to generate samples from a
multivariate distribution of \(d\) dependent variables:

\begin{itemize}
\item
  Estimate best-fit distributions \(F_1,\ldots,F_d\) for each of the
  \(d\) parameters \(X_1,\ldots,X_d\) given the lower and upper limits
  of the estimated confidence intervals for each parameter.
\item
  Sample \((z_1,\ldots,z_d)\) from a multivariate normal distribution
  \(\mathcal{N}(\mathbf{0},\Sigma)\) where the variances in \(\Sigma\)
  are all 1.
\item
  Since the marginals of this normal distribution are all
  \(\mathcal{N}(0,1)\), compute \(u_i=\Phi(z_i)\) where \(\Phi\) is the
  cumulative distribution function of the standard normal.
\item
  Finally, for each \(i=1,\ldots,d\), compute \(x_i=F^{-1}_i(u_i)\)
  where \(F_i\) is the best-fit marginal distribution of parameter
  \(i\).
\end{itemize}

The resulting vector \((x_1,\ldots,x_d)\) will be a sample from the
multivariate distribution of \((X_1,\ldots,X_d)\). Note that the
dependence structure was completely specified through the covariance
matrix \(\Sigma\) (since the covariances are assumed to be 1, this
really is a correlation matrix) and marginal distributions for each
parameter were specified by \(F_i, i=1,\ldots,d\).

\hypertarget{results}{%
\section*{Results}\label{results}}
\addcontentsline{toc}{section}{Results}

We repeat the 2 examples from Henrion (2021) here, but look at the
effect of specifying a dependence between the input parameters.

\hypertarget{hdv-prevalence-in-the-general-population}{%
\subsection*{1. HDV prevalence in the general
population}\label{hdv-prevalence-in-the-general-population}}
\addcontentsline{toc}{subsection}{1. HDV prevalence in the general
population}

With an application to hepatitis D and B viruses (HDV and HBV
respectively) from Stockdale et al. (2020), Henrion (2021) showed how to
use \texttt{bootComb} to obtain a valid confidence interval for
\(\hat{p}_{aHDV}\), the prevalence of HDV specific immunoglobulin G
antibodies (anti-HDV) in the general population.

HBV is a pre-condition for HDV and hence to derive \(\hat{p}_{aHDV}\)
Stockdale et al. (2020), obtained estimates of the prevalence of surface
antigen of the hepatitis B virus (HBsAg), \(\hat{p}_{HBsAg}=3.5\%\), and
the conditional prevalence of anti-HDV given the presence of HBsAg,
\(\hat{p}_{aHDV|HBsAg}=4.5\%\):

\begin{itemize}
\tightlist
\item
  \(\hat{p}_{HBsAg}=3.5\%\) with 95\% CI \((2.7\%,5.0\%)\).
\item
  \(\hat{p}_{aHDV|HBsAg}=4.5\%\) with 95\% CI \((3.6\%,5.7\%)\).
\end{itemize}

Assuming these 2 parameters to be independent, Henrion (2021) derived a
95\% confidence interval for the estimate
\(\hat{p}_{aHDV}=\hat{p}_{aHDV|HBsAg}\cdot\hat{p}_{HBsAg}\) using
\texttt{bootComb}, \((0.11\%,0.25\%)\).

If, however, the 2 input prevalences are not independent, e.g.~if
anti-HDV is more common among people with presence of HBsAg the higher
the population prevalence of HBsAg is, then that assumption of
independence would not hold. We can investigate how strong an effect
dependence of the parameters can have on the resulting confidence
estimate. For example, let's run the same example using
\texttt{bootComb} with specifying the following covariance matrix for
the bivariate normal copula:

\[
\Sigma=\left(
\begin{matrix}
1 & 0.5 \\
0.5 &  1
\end{matrix}
\right)
\]

\begin{Shaded}
\begin{Highlighting}[]
\FunctionTok{library}\NormalTok{(bootComb)}

\NormalTok{combFunEx}\OtherTok{\textless{}{-}}\ControlFlowTok{function}\NormalTok{(pars)\{pars[[}\DecValTok{1}\NormalTok{]]}\SpecialCharTok{*}\NormalTok{pars[[}\DecValTok{2}\NormalTok{]]\}}
\FunctionTok{bootComb}\NormalTok{(}\AttributeTok{distributions=}\FunctionTok{c}\NormalTok{(}\StringTok{"beta"}\NormalTok{,}\StringTok{"beta"}\NormalTok{),}
         \AttributeTok{qLowVect=}\FunctionTok{c}\NormalTok{(}\FloatTok{0.027}\NormalTok{,}\FloatTok{0.036}\NormalTok{),}
         \AttributeTok{qUppVect=}\FunctionTok{c}\NormalTok{(}\FloatTok{0.050}\NormalTok{,}\FloatTok{0.057}\NormalTok{),}
         \AttributeTok{combFun=}\NormalTok{combFunEx,}
         \AttributeTok{Sigma=}\FunctionTok{matrix}\NormalTok{(}\AttributeTok{byrow=}\ConstantTok{TRUE}\NormalTok{,}\AttributeTok{ncol=}\DecValTok{2}\NormalTok{,}\FunctionTok{c}\NormalTok{(}\DecValTok{1}\NormalTok{,}\FloatTok{0.5}\NormalTok{,}\FloatTok{0.5}\NormalTok{,}\DecValTok{1}\NormalTok{)),}
         \AttributeTok{doPlot=}\ConstantTok{TRUE}\NormalTok{,}
         \AttributeTok{method=}\StringTok{"hdi"}\NormalTok{,}
         \AttributeTok{N=}\FloatTok{1e6}\NormalTok{,}
         \AttributeTok{seed=}\DecValTok{123}\NormalTok{)}
\end{Highlighting}
\end{Shaded}

This yields the 95\% confidence interval \((0.10\%,0.26\%)\), a slightly
wider interval -- which makes sense, as the positive correlation means
it is more likely for pairs of bootstrapped input parameters to be both
near the upper (respectively lower) end of their confidence intervals.

For this particular application, a dependence between both prevalence
parameters, \(\hat{p}_{HBsAg}\) and \(\hat{p}_{aHDV|HBsAg}\), is
unlikely and we have therefore not considered this example any further.

\hypertarget{sars-cov-2-seroprevalence-adjusted-for-test-sensitivity-and-specificity}{%
\subsection*{2. SARS-CoV-2 seroprevalence adjusted for test sensitivity
and
specificity}\label{sars-cov-2-seroprevalence-adjusted-for-test-sensitivity-and-specificity}}
\addcontentsline{toc}{subsection}{2. SARS-CoV-2 seroprevalence adjusted
for test sensitivity and specificity}

Henrion (2021) gave an example of adjusting an estimated SARS-CoV-2
seroprevalence for the estimated sensitivity and specificity of the test
assay. Specifically:

\begin{itemize}
\tightlist
\item
  84 out of 500 study participants tested positive for SARS-CoV-2
  antibodies, yielding a seroprevalence estimate
  \(\hat{\pi}_{raw}=16.8\%\) with exact binomial 95\% CI
  \((13.6\%,20.4\%)\).
\item
  Estimated assay sensitivity: 238 out of 270 known positive samples
  tested positive \(\hat{p}_{sens}=88.1\%\), 95\% CI
  \((83.7\%,91.8\%)\).
\item
  Estimated assay specificity: 82 out of 88 known negative samples
  tested negative \(\hat{p}_{spec}=93.2\%\), 95\% CI
  \((85.7\%,97.5\%)\).
\end{itemize}

Assuming the sensitivity and specificity to be independent, Henrion
(2021) reported an adjusted seroprevalence estimate \(\hat{\pi}=12.3\%\)
with 95\% CI \((3.9\%,19.0\%)\).

However in this case, the assumption of independence is not fully
realistic: there is a trade-off between sensitivity and specificity of
the test assay, and as such one would expect a negative dependence
between the two parameters: sensitivity can be increased at the cost of
decreased specificity and vice versa.

Assuming that the sensitivity and specificity are negatively correlated
with the copula correlation parameter \(\rho=-0.5\) between these two
parameters, using the extension of \texttt{bootComb} we can now account
for the dependence of the parameters:

\begin{Shaded}
\begin{Highlighting}[]
\FunctionTok{adjPrevSensSpecCI}\NormalTok{(}
    \AttributeTok{prevCI=}\FunctionTok{c}\NormalTok{(}\FloatTok{0.136}\NormalTok{,}\FloatTok{0.204}\NormalTok{),}
    \AttributeTok{sensCI=}\FunctionTok{c}\NormalTok{(}\FloatTok{0.837}\NormalTok{,}\FloatTok{0.918}\NormalTok{),}
    \AttributeTok{specCI=}\FunctionTok{c}\NormalTok{(}\FloatTok{0.857}\NormalTok{,}\FloatTok{0.975}\NormalTok{),}
    \AttributeTok{Sigma=}\FunctionTok{matrix}\NormalTok{(}\AttributeTok{byrow=}\ConstantTok{TRUE}\NormalTok{,}\AttributeTok{ncol=}\DecValTok{3}\NormalTok{,}\FunctionTok{c}\NormalTok{(}\DecValTok{1}\NormalTok{,}\DecValTok{0}\NormalTok{,}\DecValTok{0}\NormalTok{,}\DecValTok{0}\NormalTok{,}\DecValTok{1}\NormalTok{,}\SpecialCharTok{{-}}\FloatTok{0.5}\NormalTok{,}\DecValTok{0}\NormalTok{,}\SpecialCharTok{{-}}\FloatTok{0.5}\NormalTok{,}\DecValTok{1}\NormalTok{)),}
    \AttributeTok{doPlot=}\ConstantTok{TRUE}\NormalTok{,}
    \AttributeTok{prev=}\DecValTok{84}\SpecialCharTok{/}\DecValTok{500}\NormalTok{,}
    \AttributeTok{sens=}\DecValTok{238}\SpecialCharTok{/}\DecValTok{270}\NormalTok{,}
    \AttributeTok{spec=}\DecValTok{82}\SpecialCharTok{/}\DecValTok{88}\NormalTok{,}
    \AttributeTok{seed=}\DecValTok{123}\NormalTok{)}
\end{Highlighting}
\end{Shaded}

The reported confidence interval is now \((3.8\%,19.4\%)\) - marginally
wider than when the dependence was ignored.

If we additionally specify \texttt{returnBootVals=TRUE} in the function
call, we can extract and plot the sampled pairs of sensitivity and
specificity values to check the dependence structure. This is shown on
Figure @ref(fig:Fig1): as the correlation parameter \(\rho\) in the
copula between the sensitivity and specificity is decreased from 0 to
-1, the dependence between both parameters becomes more and more
pronounced as one would expect.

This shows that a simple correlation matrix specified for the Gaussian
copula results in this case in a non-trivial dependence structure
between two beta-distributed variables, respecting the specified
marginal distributions.

We can also visualise the effect on the estimated confidence interval,
as shown on Figure Figure @ref(fig:Fig2). We can see that in this case,
with a negative correlation, the width of the CI increases at the
correlation becomes stronger. However, looking at the scale of the
y-axis we see that this is just a marginal effect.

\hypertarget{conclusions}{%
\section*{Conclusions}\label{conclusions}}
\addcontentsline{toc}{section}{Conclusions}

The R package \texttt{bootComb} has been extended and, using Gaussian
copulas, it can now handle the case of dependent input parameters. For
many applications, the effect of dependence between the parameters will
be marginal or even negligible. However, the package now allows users to
do sensitivity analyses to assess the effects of a miss-specified
dependence structure between the parameters that are being combined.

At the time of publication, the most recent version of \texttt{bootComb}
was 1.1.2.

\hypertarget{figures}{%
\section*{Figures}\label{figures}}
\addcontentsline{toc}{section}{Figures}

\begin{figure}
\centering
\includegraphics{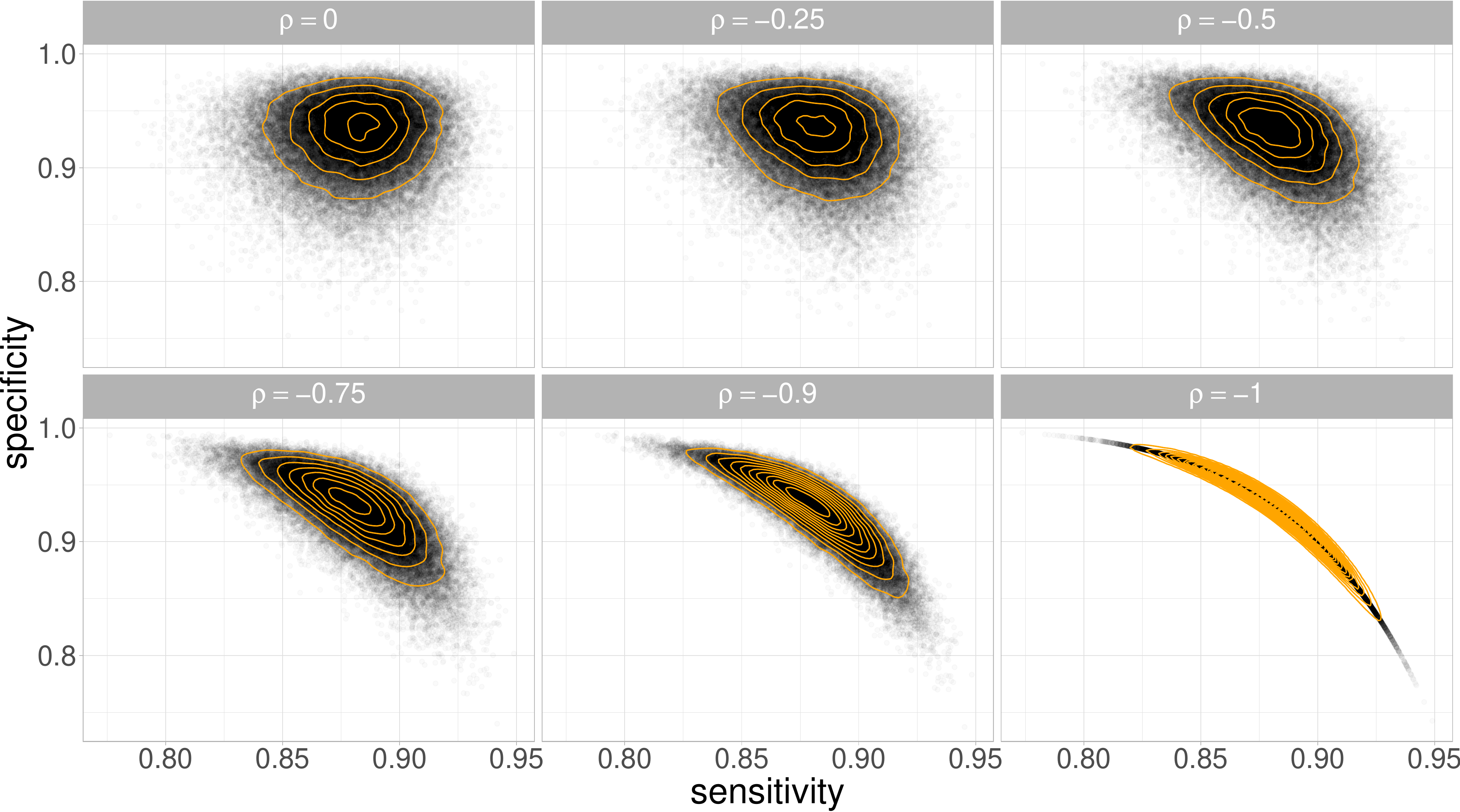}
\caption{Scatterplots showing the bootstrapped values of sensitivity and
specificity for different strenghts of dependence (from independence to
perfect correlation) between sensitivity and specifity. The empirical
kernel density estimate for the bivariate distribution in each case is
shown as orange contour lines.}
\end{figure}

\begin{figure}
\centering
\includegraphics{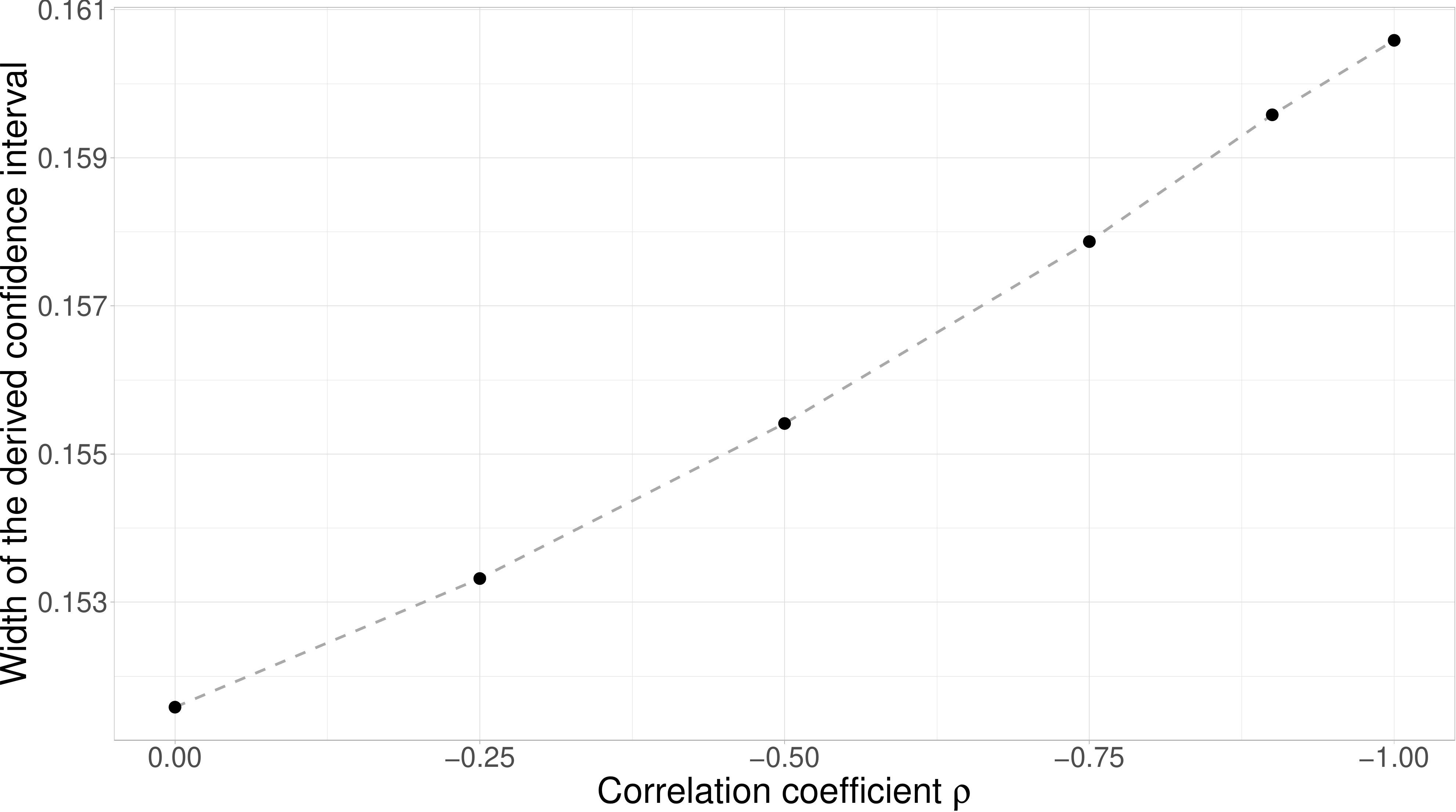}
\caption{Width of the estimated confidence interval as a function of
inreased strength of the negative correlation between sensitivity and
specificity.}
\end{figure}

\hypertarget{funding-information-see-funding-information-section-for-more-information}{%
\section{Funding Information (see funding information section for more
information)}\label{funding-information-see-funding-information-section-for-more-information}}

This research was funded in whole, or in part, by the Wellcome Trust
{[}grant: 206545/Z/17/Z{]}. For the purpose of open access, the author
has applied a CC BY public copyright licence to any Author Accepted
Manuscript version arising from this submission.

\hypertarget{data-availability-statement}{%
\section*{Data Availability
Statement}\label{data-availability-statement}}
\addcontentsline{toc}{section}{Data Availability Statement}

All data to support this work are contained within the article. The
software package itself is available from
\url{https://cran.r-project.org/package=bootComb}.

\hypertarget{conflicts-of-interest}{%
\section*{Conflicts of interest}\label{conflicts-of-interest}}
\addcontentsline{toc}{section}{Conflicts of interest}

Author Marc Y. R. Henrion declares none.

\hypertarget{references}{%
\section*{References}\label{references}}
\addcontentsline{toc}{section}{References}

\hypertarget{refs}{}
\begin{CSLReferences}{1}{0}
\leavevmode\vadjust pre{\hypertarget{ref-angus1994}{}}%
Angus, J. E. (1994). The Probability Integral Transform and Related
Results. \emph{SIAM Review}, \emph{36}(4), 652--654.
\url{http://www.jstor.org/stable/2132726}

\leavevmode\vadjust pre{\hypertarget{ref-henrion2022}{}}%
Henrion, M. Y. (2022). \emph{bootComb: Combine Parameter Estimates via
Parametric Bootstrap} (R package version 1.1.2) {[}Computer software{]}.
\url{https://cran.r-project.org/package=bootComb}

\leavevmode\vadjust pre{\hypertarget{ref-henrion2021}{}}%
Henrion, M. Y. (2021). bootComb---an R package to derive confidence
intervals for combinations of independent parameter estimates.
\emph{International Journal of Epidemiology}, \emph{50}(4), 1071--1076.
\url{https://doi.org/10.1093/ije/dyab049}

\leavevmode\vadjust pre{\hypertarget{ref-mandoloetal2021}{}}%
Mandolo, J. J., Henrion, M. Y. R., Mhango, C., Chinyama, E., Wachepa,
R., Kanjerwa, O., Malamba-Banda, C., Shawa, I. T., Hungerford, D.,
Kamng'ona, A. W., Iturriza-Gomara, M., Cunliffe, N. A., \& Jere, K. C.
(2021). Reduction in Severity of All-Cause Gastroenteritis Requiring
Hospitalisation in Children Vaccinated against Rotavirus in Malawi.
\emph{Viruses}, \emph{13}(12), 2491.
\url{https://doi.org/10.3390/v13122491}

\leavevmode\vadjust pre{\hypertarget{ref-rcoreteam2021}{}}%
R Core Team. (2021). \emph{R: A language and environment for statistical
computing}. R Foundation for Statistical Computing.
\url{https://www.R-project.org/}

\leavevmode\vadjust pre{\hypertarget{ref-sklar1959}{}}%
Sklar, M. (1959). \emph{Fonctions de Répartition À N Dimensions Et Leurs
Marges} (Issue 8, pp. 229--231). Publications de l'Institut Statistique
de l'Université de Paris.

\leavevmode\vadjust pre{\hypertarget{ref-stockdaleetal2020}{}}%
Stockdale, A. J., Kreuels, B., Henrion, M. Y. R., Giorgi, E.,
Kyomuhangi, I., de Martel, C., Hutin, Y., \& Geretti, A. M. (2020). The
global prevalence of hepatitis D virus infection: Systematic review and
meta-analysis. \emph{Journal of Hepatology}, S0168827820302208.
\url{https://doi.org/10.1016/j.jhep.2020.04.008}

\end{CSLReferences}

\end{document}